\documentclass[fleqn,usenatbib]{mnras}

\usepackage{newtxtext,newtxmath}
\usepackage[T1]{fontenc}

\DeclareRobustCommand{\VAN}[3]{#2}
\let\VANthebibliography\thebibliography
\def\thebibliography{\DeclareRobustCommand{\VAN}[3]{##3}\VANthebibliography}

\usepackage{graphicx}	
\usepackage{amsmath}	
\newcommand{\bfbeta}{\boldsymbol{\beta}}
\newcommand{\bfalpha}{\boldsymbol{\alpha}}
\newcommand{\bftheta}{\boldsymbol{\theta}}
\newcommand{\Dl}{D_{\rm l}}
\newcommand{\zl}{z_{\rm l}}
\newcommand{\Dls}{D_{\rm ls}}
\newcommand{\Ds}{D_{\rm s}}
\newcommand{\fsrc}{f_{\rm s}}
\newcommand{\fsrcmin}{\fsrc^{\rm min}}
\newcommand{\ts}{t_{\rm s}}
\newcommand{\td}{t_{\rm d}}

\newcommand{\Eobs}{E}
\newcommand{\trest}{\tilde t}
\newcommand{\Erest}{\tilde E}
\newcommand{\swiftbat}{ \emph{Swift}/BAT }
\newcommand{\Arest}{\tilde A}
\newcommand{\Epeakrest}{\tilde E_{\rm peak}}
\newcommand{\Ezerorest}{\tilde E_{0}}
\newcommand{\Epeak}{E_{\rm peak}}
\newcommand{\Enorm}{E_{\rm norm}}

\newcommand{\fbg}{f_{\rm bg}}
\newcommand{\Omegabg}{{\cal A}_{\rm bg}}
\newcommand{\Omegad}{{\cal A}_{\rm d}}

\newcommand{\Emax}{E_{\rm max}}

\newcommand{\keV}{{\rm keV}}
\newcommand{\Emin}{E_{\rm min}}

\newcommand{\Dlum}{D_{\rm lum}}
\newcommand{\mdm}{m_{\rm dm}}

\title[Primordial black hole abundance using lensing parallax]{On the feasibility of primordial black hole abundance constraints using lensing parallax of GRBs}

\author[Priyanka Gawade et al.]{
Priyanka Gawade,$^{1}$\thanks{E-mail: priyankag@iucaa.in}
Surhud More$^{1,2}$\thanks{E-mail: surhud@iucaa.in},
Varun Bhalerao$^{3}$
\\
$^{1}$Inter-University Centre for Astronomy and Astrophysics, Ganeshkhind, Pune 411007, IN\\
$^{2}$Kavli Institute for the Physics and Mathematics of the Universe (WPI), University of Tokyo, 5-1-5, Kashiwanoha, 2778583, JP\\
$^{3}$Department of Physics, Indian Institute of Technology Bombay, Powai, Mumbai, 400076, IN \\
}

\date{Accepted XXX. Received YYY; in original form ZZZ}

\pubyear{2015}

\begin{document}
\label{firstpage}
\pagerange{\pageref{firstpage}--\pageref{lastpage}}
\maketitle

\begin{abstract}
Primordial black holes, which could have formed during the early Universe through overdensities in primordial density fluctuations during inflation, are potential candidates for dark matter. We explore the use of lensing parallax of Gamma ray bursts (GRBs), which results in different fluxes being observed from two different vantage points, in order to probe the abundance of primordial black holes in the unexplored window within the mass range $[10^{-15}-10^{-11}]M_\odot$. We derive the optical depth for the lensing of GRBs with a distribution of source properties and realistic detector sensitivities. We comment on the ability of the proposed Indian twin satellite mission Daksha in its low earth orbit to conduct this experiment. If the two Daksha satellites observe 10000 GRBs simultaneously and the entirety of dark matter is made up of $[10^{-15}-10^{-12}]M_\odot$ black holes, Daksha will detect non-zero lensing events with a probability ranging from 80 to 50 per cent at the bin edges, respectively. Non-detections will not conclusively rule out primordial black holes as dark matter in this mass range. However, we show that meaningful constraints can be obtained in such a case if the two satellites are separated by at least the Earth-Moon distance.
\end{abstract}


\section{Introduction}

Astrophysical observations of the cosmic microwave background as well as that of galaxy clusters require that dark matter should account for around 85 per cent of the total matter content in the Universe, and contribute to 27 per cent of the total mass-energy-density of the Universe \citep{Planck2020}. Despite a plethora of evidence for its existence, the nature of dark matter remains an important unresolved mystery in physics. Observational evidence indicates that dark matter is non-baryonic, non-relativistic, and interacts with ordinary matter only via gravity \citep[see e.g.,][]{davis1985evolution, clowe2006direct, dodelson2006can}. The search for the particle physics candidate for dark matter has not yet yielded conclusive results. Proposed candidates for dark matter include Weakly Interacting Massive Particles \citep[WIMPs, ][]{jungman1996supersymmetric} which have now been constrained by various direct and indirect detection experiments \citep[see][ for a review]{hooper:2018}. 

Primordial black holes (PBHs) could form in the early Universe due to the presence of large overdensities and are also potential candidates for dark matter \citep{zel1967hypothesis, hawking1971gravitationally, carr1974black}. Cosmological observations including the cosmic microwave background, as well as various microlensing surveys carried out from both ground and space have ruled out a vast parameter range for the possible masses of such primordial black holes \citep[see e.g.,][ for reviews]{Sasaki:2018,2021carrkohri,carr2021pbh}. However, there exists a window in the mass range $[10^{-16}-10^{-11}]M_\odot$ where such PBHs remains unexplored thus far. Such PBHs could even make up the entirety of the dark matter, thus solving the dark matter puzzle without recourse to any new particle physics. As dark matter is known to interact only via gravity, gravitational lensing is a very useful tool to probe dark matter.

The existence of neutron stars and white dwarfs in globular clusters have been previously suggested to constrain the PBH abundances in this mass range \citep{Capela:2013a, Capela:2013b}. However, such constraints require globular clusters to contain dark matter, the validity of its existence or its requirement in order to form globular clusters is currently inconclusive \citep{Conroy:2011, Ibata:2013}.

The PBHs in this mass range cannot be explored with light in the optical wavelength range, as the Schwarzschild radii of such black holes are smaller than the wavelength of light \citep[see e.g.][]{Niikura2019}. We need to observe sources such as gamma ray bursts (GRBs) that contain wavelengths suitable to probe the aforementioned PBH mass range. The femto-lensing of GRBs has been proposed previously to constrain the abundance of black holes in such range by searching for interference fringes in the energy spectrum of GRBs \citep{Gould:1992}. \citet{Barnacka:2012} used the non-detection of such interference in the observations of the energy spectra of GRBs seen by Fermi to claim that PBH with masses within the range $[10^{-16}, 10^{-14}] M_\odot$ cannot contribute more than $10$ per cent of dark matter. However, when such observations have been reinterpreted taking into account the finite source size of GRBs, \citet{Katz:2018} show that the non-detection of interference effects in the Fermi GRB spectra do not meaningfully constrain the abundance of PBHs.

In their prompt phase, GRBs generally last no longer than a few seconds, making it challenging to observe and analyse the light curve. However, one could observe gravitational lensing parallax resulting from two spatially separated observers simultaneously observing the same GRB that gets lensed by a PBH in between \citep[see e.g., ][]{Nemiroff:1995, Nemiroff:1998}. The magnification of a source caused by gravitational lensing depends upon the on-sky separation between the lens and the source as seen by the observer. Thus the two observers will register a different flux for the same source. This effect is called gravitational lensing parallax and can be used to constrain the PBH abundance in the aforementioned mass window. Recently, \citet{jung2020gamma}, studied the feasibility of such constraints with observers separated by a variety of distances.

Daksha \citep{bhalerao2022science} is a proposed Indian twin satellite GRB monitoring mission, well equipped to detect GRBs. The two identical Daksha satellites having suitable energy band coverage will enable a unique lensing parallax experiment with the detectors with same sensitivity and energy range for the first time, where a GRB can be observed with two different lines-of-sight. The light from the GRB will arrive at a different impact parameter compared to the PBH for the two different satellites, thus causing a difference in the measured flux of the GRB. A null detection of a difference in flux (i.e. no detection of lensed GRBs) could potentially constrain the PBH abundance in the unexplored mass range. Alternatively, a detection of a difference in the detected fluxes of the same GRB from the two satellites will open an exciting possibility of finding such tiny primordial black holes for the very first time. This experiment will require excellent cross-calibration between the detectors on board the Daksha satellites.

The analysis performed in \citet{jung2020gamma} suggests that such a twin satellite mission should be able to constrain the primordial black hole abundance. However, their analysis relies on the assumption that detectors are sensitive to fractional changes in magnification, with no limit on the absolute sensitivity. A real life detector is expected to have a sensitivity threshold which depends on various factors such as the detector area, background noise, signal-to-noise ratio (SNR) and the time duration of the GRB. We carefully account for detector characteristics in our computation, and assess whether such lensing parallax is detectable with satellites in Lower Earth Orbit (LEO). This requires the use of additional parameters of the GRB such as its unlensed flux along with detector specifications. As a result our analysis is more realistic, and can be directly applied to any twin satellite mission once its parameters are known.

This paper is organized as follows. We present a brief introduction to gravitational lensing and lensing parallax in Sec.~\ref{sec:lensing}. We derive the realistic conditions for the detection of lensing parallax in Sec.~\ref{sec:detector_conditions} and present the distribution of the observed GRB parameters that we used for our lensing calculation in Sec.~\ref{sec:grbdist}. We present our methodology to compute the optical depth in Sec.~\ref{sec:tau}, and present and discuss our results in Sec.~\ref{sec:results}. Finally we summarize our findings in Sec.~\ref{sec:summary}. Throughout this paper, we use an overhead $\tilde .$ on symbols to denote quantities observed in the rest frame while symbols without overheads are in the observed frame.

\section{Gravitational lensing}
\label{sec:lensing}

\subsection{Lensing due to point mass}
The lens equation relates the true position of a point source ($\bfbeta$) as seen in the plane of the sky, and the resultant image position ($\bftheta$) via the deflection angle ($\bfalpha$),
\begin{align}
    \bfbeta = \bftheta - \bfalpha(\bftheta)
\end{align}
The scaled deflection angle of light due to a point mass $M$ at an angle $\bftheta$ away from the point mass is given by
\begin{align}
\bfalpha = \frac{4GM}{c^2\Dl} \frac{\bftheta}{|\bftheta|^2}\,,
\end{align}
where $\Dl$ is the angular diameter distance to the lens. The Einstein angle, $\theta_{\rm E}$ corresponds to the angular radius of the Einstein ring that forms when the source is perfectly aligned with the lens location ($\bfbeta = 0$),
\begin{align}
\label{eq:Einstein_ang}
\theta_{\rm E} = \sqrt{\frac{4GM}{c^2}\frac{\Dls}{\Dl\Ds}}\,,
\end{align}
where $\Ds$ and $\Dls$ correspond to the angular diameter distance of source to us, and between the lens and the source, respectively. The Einstein angle can be used to scale the true angular position of the source and the image with respect to the lens centre. Due to the azimuthal symmetry the lens equation can be written as
\begin{align}
    y = x - \frac{1}{x}\,,
\end{align}
where $y=\beta/\theta_{\rm E}$, and $x=\theta/\theta_{\rm E}$. For a given value of the source location $y$, there are two images which form at
\begin{align}
   x = \frac{y \pm \sqrt{y^2+4}}{2}\,,
\end{align}
with magnifications given by
\begin{align}
    \mu_{\pm} = \left| \frac{1}{2} \pm \frac{y^2 + 2}{2y\sqrt{y^2 + 4 }}\right | \,,
\end{align}
which can reach arbitrarily large values for near perfect alignment ($y\rightarrow 0$).

For lensing objects with small values of $\theta_{\rm E}$, the multiple images are not well separated. In this case, and in the limit of geometrical optics, the resultant total observed magnification is equal to the sum of the two magnifications above $\mu = |\mu_{+}| + |\mu_{-}| $. If the wavelength of light used to observe the images is $\lambda \gtrapprox 4\pi R_{\rm s}$, where $R_{\rm s}$ denotes the Schwarzschild radius, one needs to additionally consider effects due to wave optics. The magnification starts to reduce for wavelengths larger than this. For high-energy sources observed at an energy range in $[20\,{\rm keV}, 200\,{\rm keV}]$, the geometrical optics approach is valid for masses above $ 10^{-15}M_{\odot}$ (see Appendix~\ref{app:GOL}).

The above expressions for magnification have assumed that the source can be approximated as a point, with an angular size much smaller than the relevant angular scale in the problem, the Einstein radius. In case this assumption does not hold, then the above equations get modified as each part of the source gets magnified in a differential manner. A fitting function to describe the magnification for such a case has been presented in \citet[][see Appendix~\ref{app:finite}]{1994ApJ...430..505W} in terms of $\delta$, the angular size of the source scaled by the Einstein angle, and the true position of the source ($y$). The finite source size causes the magnification to saturate to a value given by $\sqrt{1+4\delta^{-2}}$, when the source is perfectly aligned.

\subsection{Lensing parallax}

When the same lensing system is observed from two locations $D_1$ and $D_2$, the true source positions with respect to the lens will be different due to parallax. Considering the plane which passes through $D_1, D_2$ and the source $G$ (see Fig.~\ref{fig:example}), the parallax of the lens arises due to the angle differences within this plane, while the out of plane component is the same for the two positions (see Appendix ~\ref{app:parallax}). The angular difference, $\Pi$, is given by
\begin{align}
    \Pi = \frac{R_{\rm sep}\sin\theta}{\Ds}\frac{\Dls}{(1+\zl)\Dl \theta_{\rm E}} \,,
    \label{eq:parallax}
\end{align}
where $R_{\rm sep}$ is the separation between $D_1$ and $D_2$, and $\theta$ is the angle between the line joining the two satellites, $D_1D_2$, and the line joining its midpoint to the source position $G$.

The difference in the source positions as seen from $D_1$ and $D_2$ will result in different magnification factors as seen by observers at these locations, which can be detected via the measured flux to infer the presence of an intervening point mass in between these positions and the source. If we consider the two observing locations $D_1$ and $D_2$ to be low Earth satellites which are located diametrically opposite to each other on identical orbits, the projected separation will vary for a given line-of-sight to the source. When computing the probability of lensing, we will need to average over all possible orbital positions as well as source positions. However, the statistical average over all possible positions for the source for a given orbital position renders the averaging over the entire orbit of the satellites unnecessary.

\section{Conditions for the detection of lensing parallax}
\label{sec:detector_conditions}

\begin{figure}
    \includegraphics[scale=0.5]{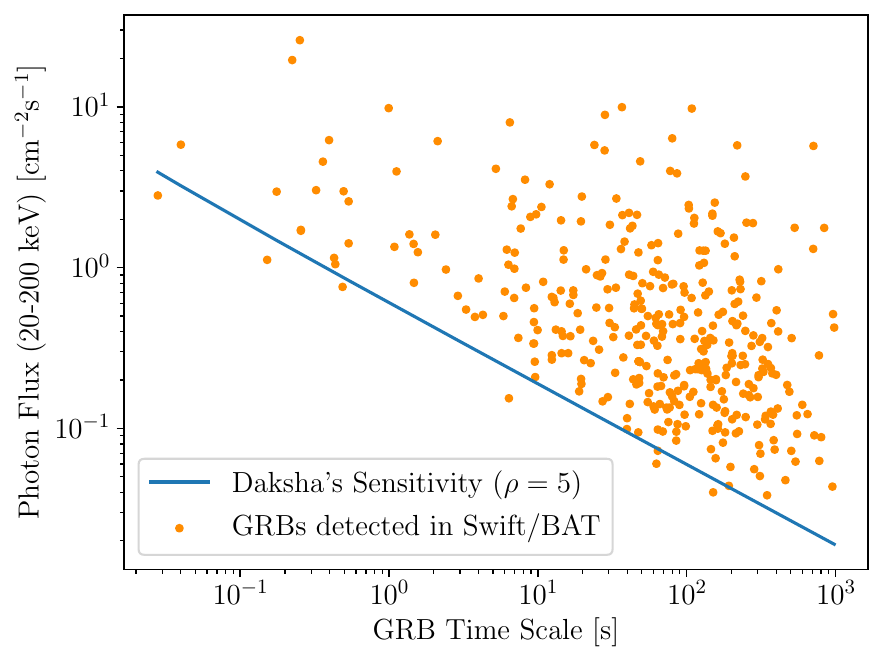}\caption {Photon flux in Daksha's energy band for GRBs detected by \swiftbat are shown as points with orange colour. These can be compared to Daksha's sensitivity for detecting GRBs at a SNR of 5 (shown with a blue line) for a GRB that lasts for a given time scale.} \label{fig:sensitivity_time}
\end{figure}

Consider a gamma ray source with photon flux $\fsrc$, which lasts for a duration $\ts$. The gamma ray detector panels will be arranged in a certain geometry. The effective area for the detection of the flux from the source, $\Omegad$, will be the area of these detectors projected perpendicular to the line-of-sight to the source. The background noise will however be proportional to the total area of the detectors $\Omegabg$. Let the background noise be given by $\fbg$. We assume that the background count rate is well-measured from longer pre-GRB and post-GRB intervals, or from other detectors on the spacecraft. The signal to noise ratio $\rho$ is given by
\begin{align} \label{eq:snr}
\rho=\frac{\fsrc \Omegad \ts}{\sqrt{\fsrc \Omegad \ts+ \fbg \Omegabg \ts}}=\frac{\fsrc \Omegad}{\sqrt{\fsrc \Omegad + \fbg \Omegabg }}\sqrt{\ts}\,.
\end{align}
We can invert this equation to calculate the minimum flux of source $\fsrc$, received in a time interval $\ts$ which can be detected above a signal to noise ratio $\rho$, 
\begin{align} \label{eq:fsrcmin}
\fsrcmin =\frac{\rho^2+\sqrt{\rho^4+4 \rho^2 \fbg \Omegabg \ts}}{2 \Omegad \ts}\,.
\end{align}
We show this flux limit for the high sensitivity medium energy band detectors on Daksha as a function of $\ts$ with the detector parameters $\Omegad = 1300\,{\rm cm}^2$, $\Omegabg = 2400\,{\rm cm}^2$, $\fbg = 10\,{\rm ph}\,{\rm cm}^{-2}\,s^{-1}$ and for a signal to noise threshold of $\rho = 5$ as solid blue line in Fig.~\ref{fig:sensitivity_time}. As expected, we see that the fainter sources can be detected if they emit for a longer period of time with a flux limit that falls approximately proportional to $\sqrt{\ts}$. In the same figure, we also show the GRBs that were detected by \swiftbat with orange points. Given the very similar sensitivities of Daksha and \swiftbat, we observe that our calculated flux limits as a function of $\ts$ forms the lower envelope of the detected GRB population from \swiftbat.

We assume that the energy spectrum of gamma ray sources is given by the Band function in the rest frame of the GRB \citep{1993ApJ...413..281B},
\begin{align}
\frac{dN}{d\Erest d\trest} = \Arest \begin{cases}
			\left(\frac{\Erest}{100 \keV}\right)^\alpha {\rm e}^{\left(-\frac{\Erest}{\Ezerorest}\right)}, & \frac{\Erest}{\alpha-\beta}\leq  \Ezerorest  \\
            \left[\frac{(\alpha-\beta) \Ezerorest}{100 \keV}\right]^{\alpha-\beta} {\rm e}^{\beta-\alpha}\left(\frac{\Erest}{100 \keV}\right)^\beta, & \frac{\Erest}{\alpha-\beta}\geq \Ezerorest \,, 
            \end{cases}
\end{align}
where $\alpha, \beta$ and $\Ezerorest$ are free parameters, where $\Ezerorest =\frac{\Epeakrest}{\alpha+2}  $ and the normalization $\Arest$ has units of number of photons per unit energy per unit time. This normalization can be computed for any GRB whose fluence ${\cal F}$ (in units of ${\rm erg\, cm}^{-2}$) and $T_{90}$ have been measured in an observed energy band $[\Emin, \Emax]$ by using \footnote{This assumes that the above energy spectrum is averaged over a time interval $T_{90}$.}
\begin{align}\label{eq:f1}
    \frac{\cal F}{T_{90}} = \frac{1}{4\pi\Dlum^2(z)}\int_{(1+z)\Emin}^{(1+z)\Emax} \Erest \frac{dN}{d\Erest d\trest} d\Erest\,,
\end{align}
where $\Dlum$ is the luminosity distance to the source. Once the normalization is obtained, the photon flux, $\fsrc$, received in an observed band $[\Emin, \Emax]$ is given by
\begin{align}\label{eq:f2}
\fsrc &= \frac{(1+z)}{4\pi \Dlum^2(z)}\int_{(1+z)\Emin}^{(1+z)\Emax} \frac{dN}{d\Erest d\trest } d\Erest\,.
\end{align}

We will require that the difference in the fluxes seen from the two satellites which observe a total magnification of $\mu_1$ and $\mu_2$ can be detected given the noise in the measurements in each of the detectors with a signal to noise ratio greater than $\rho$,
\begin{align} \label{eq:detection}
\fsrc \Omegad \ts \left(|\mu_1-\mu_2|\right) \geqslant \rho \sqrt{{\fsrc \Omegad \ts \left(\mu_1+\mu_2\right)}+2 \fbg \Omegabg \ts}\,.
\end{align}

In the description above we have assumed that the shape of energy spectrum of all GRBs is given by the same parameters. In reality we expect some diversity in the GRB parameters. In order to capture this diversity, and assess its impact, as a separate case, we also consider the entire database of spectral parameters inferred from the GRBs present in the \href{https://swift.gsfc.nasa.gov/results/batgrbcat/index_tables.html}{\swiftbat catalogue}\footnote{\url{https://swift.gsfc.nasa.gov/results/batgrbcat/index_tables.html}}. The observed energy spectrum for the GRBs in this catalogue was fit with either a power law or a cut-off power law parameterization, such that
\begin{align}
    \left.\frac{df}{d\Eobs}\right|_{\rm PL} &= A\left[\frac{E}{\Enorm}\right]^\alpha\,, \\
    \left.\frac{df}{d\Eobs}\right|_{\rm CPL} &= A\left[\frac{E}{\Enorm}\right]^\alpha \exp\left[-\left(\frac{E(2+\alpha)}{\Epeak}\right)\right]\,.
\end{align}
We used the best fit values of the observed spectral parameters for each GRB for the purpose of our calculation. In this case we expect the distribution of these parameters to be broader than the true distribution due to the presence of errors in the measurement. Nevertheless the two cases, one with average parameter values, and the other with individual parameter values for each GRB, are expected to bracket the true distribution of the parameters that describe the energy spectrum.

\section{\swiftbat GRB parameter distribution}
\label{sec:grbdist}

The equations in the above section give us a criteria for the detection of the lensing parallax from sources with unlensed photon fluxes equal to $\fsrc$, and observed from two satellites with effective magnifications, $\mu_1$ and $\mu_2$. In this paper, we will use the GRBs already detected by \swiftbat in the energy range $[15, 150]$~$\keV$, and compute the probabilities of detecting lensing parallax from a similar set of GRBs given the detectors planned on board the twin satellite mission, Daksha, in the energy range $[20, 200]$~$\keV$. As shown in Fig.~\ref{fig:sensitivity_time}, this assumption is reasonable given the similar sensitivity of the two detectors. 

\begin{figure*}
\includegraphics[width=1\textwidth]{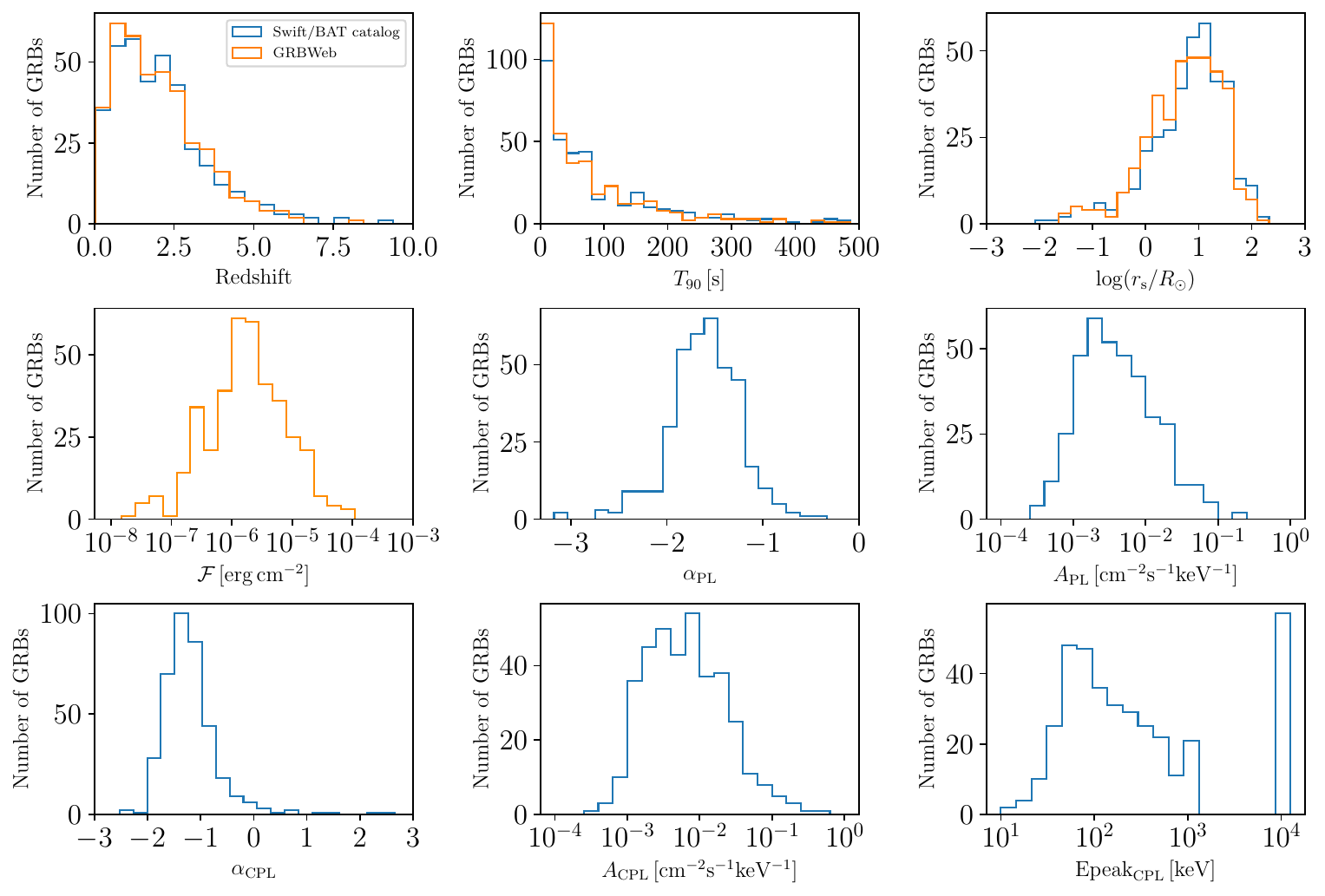}
\caption{Distribution of redshift, the timescale $T_{90}$, size ($r_{\rm s}$), fluence (${\cal F}$), together with the individual determinations of the spectral parameters corresponding to both the power law and the cut-off power law fits for GRBs. Data retrieved from the \href{https://user-web.icecube.wisc.edu/~grbweb_public/}{GRBWeb database} is shown using histograms in orange colour, while that retrieved from \href{https://swift.gsfc.nasa.gov/results/batgrbcat/index_tables.html}{\swiftbat catalogue} is shown in blue, whenever available. We use the parameters from these two databases to compute the optical depth for a detectable lensing parallax in two alternative ways.}

\label{fig:distributions}
\end{figure*}

In the first case, we use the Band function with average spectral parameters for the GRB sources detected by \swiftbat  listed at the \href{https://user-web.icecube.wisc.edu/~grbweb_public/}{GRBWeb} database\footnote{\url{https://user-web.icecube.wisc.edu/~grbweb\_public/}}. We obtain the redshift $z$, time scale $T_{90}$ and fluence, i.e. energy per unit area (in the units of ${\rm erg}/{\rm cm}^{-2}$) for these GRBs detected by \swiftbat. The distributions of these parameters are shown in Fig.~\ref{fig:distributions}. Magnification due to lensing will depend upon the ratio of the angular size of the source compared to the Einstein radius of the lens. The physical transverse emission size $r_{\rm s}$ of a GRB source is related to its minimum variability time scale ($t_{\rm var}$) and the Lorentz boost factor $\Gamma$. \citet{Barnacka_2014} empirically show that the following scaling relation holds between the physical size $r_{\rm s}$ and $T_{90}$, 
\begin{align} \label{eq:sourcesize}
r_{\rm s} \sim ct_{\rm var}\Gamma/(1 + z_{\rm s}) \sim cT_{90}/(1 + z_{\rm s})\,,
\end{align}
where $z_{\rm s}$ is the source redshift. 

We obtain 382 GRBs for which all of the parameters required for our analysis are available in the database. Based on these parameters, similar to \citet{jung2020gamma}, we also observe a size distribution such that around 10\% of GRBs have $r_ {\rm s}\leqslant r_\odot$, 3\% have $r_ {\rm s}\leqslant 0.1 r_\odot$ and remaining 90\% have  $r_ {\rm s}\geqslant r_\odot$\footnote{Note that the length scale of the Einstein radius for $5\times10^{-12}M_\odot$ point mass lens at a distance of 1 Gpc with a source at 2 Gpc is approximately equal to the solar radius (see Eq.~\ref{eq:Einstein_ang}), which makes it a convenient unit to describe source sizes.}. GRBs are observed at cosmological distances up to $z_ {\rm s}\leqslant 10$, while majority of the GRBs are found at $z_ {\rm s}= 0.5 \sim 3$.

The database provides parameters such as redshift $z$, time scale $T_{90}$ and fluence for the GRBs detected by \swiftbat in the energy range $[15, 150]$~$\keV$. Owing to the slight difference in the energy range, we need to convert this fluence into photon flux that can be measured by Daksha in the energy range $[20, 200]$~$\keV$. Assuming a shape for the spectrum in the rest frame of the GRB with average values of spectral parameters $\alpha,\beta $ and $\Epeakrest$ to be $-1.08, -2.14$ and 600 $\keV$, respectively \citep[see][]{2014ApJS..211...12G}.

In the second case, we use the \href{https://swift.gsfc.nasa.gov/results/batgrbcat/index_tables.html}{\swiftbat catalogue}. We obtained 374 GRBs from this catalogue for which all of the parameters required for our calculations are available. The distributions of these parameters are shown in Fig.~\ref{fig:distributions}. The energy spectrum for the GRBs in this catalogue was fit with either a power law or a cut-off power law. We used the best fit values of the spectrum parameters for each GRB for the purpose of our calculation.

\section{Optical Depth and constraints on the PBH-dark matter abundance}
\label{sec:tau}

We assume that the PBHs are uniformly distributed with a comoving number density given by
\begin{align}
n = \rho_{\rm crit,0}\Omega_{\rm DM}f/\mdm \,,
\label{eq11}
\end{align}
where, $f = \Omega_{\rm PBH}/\Omega_{\rm DM}$ is the PBH abundance and $\mdm$ is the mass of the primordial population of black holes. Since the lensing optical depth will peak roughly midway between the lens and the source, we expect that the dark matter clustered near the source or the observer will not contribute significantly to the optical depth, $\tau$. Even though the distribution of dark matter itself is known to be clustered and could potentially weaken the constraining power of observations given null detections \citep[see e.g.,][]{Belotsky:2019}, given the large projection length to the source, the clustering signal is expected to be a relatively small contribution \citep[see Appendix D in ][]{jung2020gamma}.

For any given GRB, the primordial black holes that lie in two narrow tubes along the lines of sight to the source as observed from each of the two satellites are expected to result in an appreciable parallax that can be detected given the sensitivity of the detectors. The expected number of PBH lenses within the total volume of these tubes is expected to be $\ll 1$ and it reflects the optical depth for lensing to that particular source. For a uniform distribution of such potential lenses, we expect the probability to observe a given number of lensing events given their individual optical depths to follow the Poisson distribution.

Consider two identical detectors with detector specifications ${\cal D}$ separated by a distance $R$, simultaneously observing a GRB source at a comoving distance $\chi_{\rm S}$. The optical depth can be written as, 
\begin{align}\label{opdep}
\tau(f, \mdm, \chi_{\rm S}, {\cal S}, {\cal D}, {\cal G}, R_{\perp}) = n(f, \mdm) V(\mdm, \chi_{\rm S}, {
\cal S}, {\cal D}, {\cal G}, R_{\perp})\,,
\end{align}
where, the source properties are represented by ${\cal S} \equiv (t_{\rm s}, r_{\rm s}, {\cal F})$, the detector specifications ${\cal D}\equiv(\Omegad$, $\Omegabg$, $\fbg$, $E_{\rm min}$, $E_{\rm max}, \rho)$ with values taken from those listed in Section~\ref{sec:detector_conditions} and GRB spectrum by  ${\cal G}\equiv(\Arest, \alpha, \beta, \Epeakrest)$ for the case of Band function, $(A, \alpha)$ for PL and $(A, \alpha, E_{\rm peak})$ for CPL with values corresponding to the individual GRBs. The symbol $R_{\perp}$ denotes the component of the detector separation perpendicular to the line-of-sight. Finally, $V$ denotes the desired comoving volume containing positions of PBH lenses that can lead to detectable lensing parallax and can be expressed as 
\begin{equation}\label{}
V(\mdm, \chi_{\rm S}, {\cal S}, {\cal D}, {\cal G}, R_{\perp}) = \int^{\chi_{\rm S}}_0 \sigma(\mdm, \chi_{\rm S}, {\cal S}, {\cal D}, {\cal G}, R_{\perp}) \ d\chi_{\rm L}\,. \label{volume}
\end{equation}
Here $\sigma$ represents the comoving cross-section effective for lensing parallax detection at a comoving distance $\chi_{\rm L}$ (see Appendix~\ref{app:parallax} for how we determine the lensing cross-section). We adopt a flat $\Lambda$CDM model with the cosmological parameters $\Omega_{\rm m}=0.315$, $\Omega_{\rm b}=0.049$ and $h= 0.673$ consistent with results from the Planck collaboration \citep{Planck2020}. For a given redshift, the source can be at any angular orientation compared to the separation of the two orbiting satellites. Hence, $R_{\perp}$ changes with the source orientation and thus the cross-section $\sigma$ needs to be averaged over all possible angular positions. Such an averaging over all possible angular positions leads to an optical depth independent of the phase of the orbit, hence we do not further average over the orbital phase of the satellites. We evaluate the lensing cross section $\sigma$ in a Monte Carlo manner (see Appendix~\ref{app:parallax}). 

For a given PBH mass, the optical depth and and the single-lensing probability $P_i$, for each GRB are directly proportional to $f$, the fraction of dark matter bound in PBHs of a given mass $\mdm$. The expected number of lensing events $N_{\mathrm {exp}} = \sum_{i=1}^N P_i(f)$. Under the asssumption that we do not detect a lensing parallax for even a single event, we can compute a bound on $f$ at 95\% confidence limit. The probability of having no detection for $N$ sources is given by $P_{\rm null}(f) = \prod^N_{i=1} (1-P_{i}(f))$, where $i$ is the source index. The $95\%$ bound on $f$ corresponds to $P_{\rm null}(f) = 0.05$. We assume that Daksha will detect 10000 GRBs over an extended mission, with distributions of the source parameters similar to shown in Fig.~\ref{fig:distributions}. The expected abundance constraints will scale inversely proportional to the actual number of GRBs observed by Daksha\footnote{Note that the instrumental background will affect the signal-to-noise ratio for the detection of the lensing parallax, but once the background values are set, the expected number of detectable events with lensing parallax scale proportionally to the total number of GRBs.}.

\section{Results and Discussion}
\label{sec:results}

We present the constraints on the abundance of PBHs for a twin satellite mission like Daksha, where the two satellites are in a low Earth orbit (LEO) for two cases. In the first case, we used the $382$ GRBs from the GRBweb database, for which we use the average spectral parameters, while in the second case we used the $374$ GRBs from the \swiftbat database which have individual but noisy spectral parameters. We compute the optical depth to every GRB considering its redshift, time period, spectrum, detector specification, the orbit of satellites and the PBH mass. We then estimate the probability for lensing parallax based on the methodology explained in the previous section.

In both the cases described above, we obtain a lensing probability for a GRB that ranges in a decreasing manner from $1.7\times 10^{-4}$ to $0.7\times 10^{-4}$ in case PBHs make up the entirety of dark matter with a monochromatic mass spectrum in the mass range $[10^{-15}-10^{-12}]M_\odot$. This implies that for 10000 GRBs observed simultaneously by Daksha in its entire life span, the probability of obtaining non-zero events ranges in a decreasing manner from 80 to 50 per cent within the range. Non-detections of lensing events on the other hand will translate into constraints on the abundance of PBHs as a fraction of dark matter.

\begin{figure}

    \includegraphics[scale=0.6]{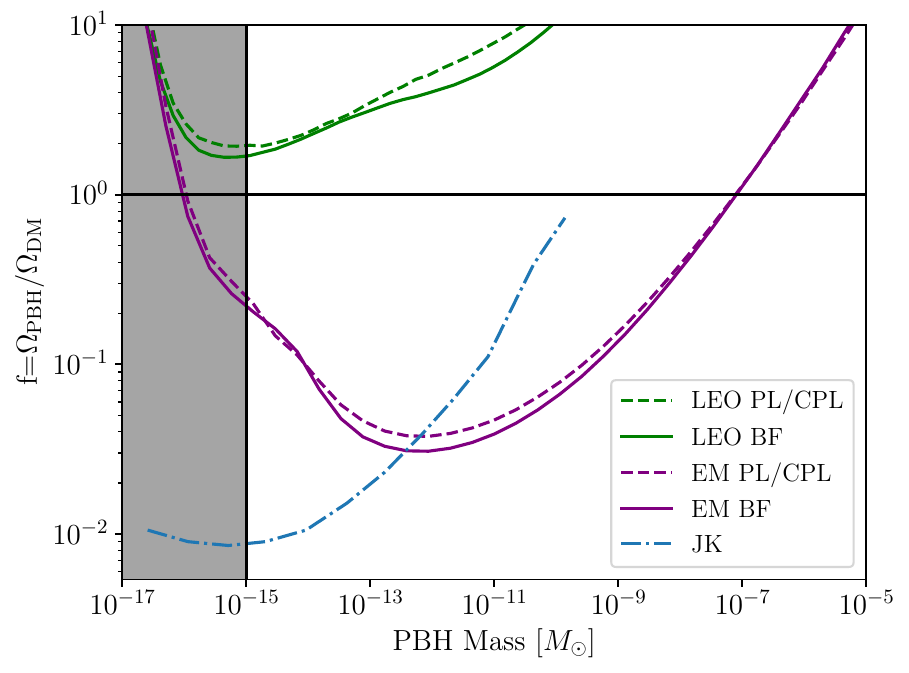} \caption{Comparison of the constraints of the PBH abundance with detectors sensitive as Daksha but with different separations (Low Earth Orbit, LEO, or Earth Moon, EM) and different choices for the spectral parameters (average Band function parameters, BF, or individual best fits from either the Power law or the Cut-off power law PL/CPL) for 10000 GRBs with the difference in fluxes observed between two detectors observed at significance larger than 5$\sigma$. The constraints predicted by \citet{jung2020gamma} for Low Earth Orbit, but with detector sensitivity that allows detection of 1 per cent differences in fluxes regardless of their fluence or duration, are shown by a dot-dashed line. The grey shaded region represents the wave optics regime, where our assumption of the geometrical optics limit is no longer valid.} \label{fig:constraints}

\end{figure}
\begin{figure}

    \includegraphics[scale=0.7]{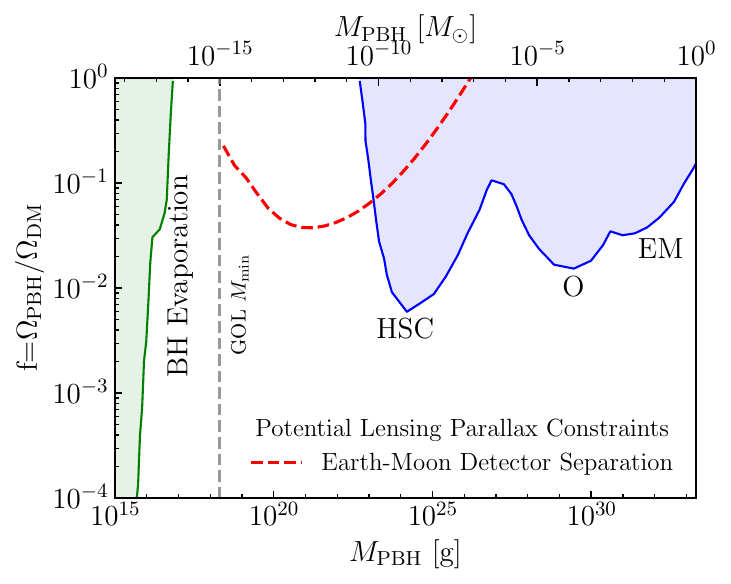} \caption{Comparison of the constraints on the PBH abundance possible with 10000 GRBs as observed by twin satellites with detectors with similar sensitivity as Daksha but separated by the distance between the Earth and the Moon (red dashed line). The grey dashed line shows the minimum mass that can be probed by Daksha conservatively in the geometrical optics limit (GOL). The green shaded region shows the bounds from PBH evaporation \citep[see e.g., ][]{carr2010, voyager, laha, derocco, integral} and the blue shaded region represents the constraints established by other microlensing experiments by Subaru \citep[HSC, ][]{Niikura2019}, EROS and MACHO collaborations \citep[EM, ][]{Gal_halo_mag_clouds, Gal_bulge_EROS2}, and OGLE 
        \citep[O, ][]{Ogle2009, Ogle2010_SMC, Ogle2011_LMC_self_lensing, Ogle2011_LMC_subsolar_machos, Ogle2011_SMC_finalconc}.} \label{fig:constraints_comp}

\end{figure}

Our results, in the case of non-detection, are summarized in Fig.~\ref{fig:constraints}. The green solid and dashed lines show the result for a Daksha-like configuration, two satellites in a LEO for the two cases that were summarized above, respectively. We see that there is not much of a difference in the expected constraints whether we use the average spectral parameters or the individual parameter values for the GRBs. More importantly we observe that given Daksha's sensitivity, a non-detection will not conclusively, (i.e. at $>95$ per cent) be able to rule out the primordial black hole mass window, since the number of expected lensing events even with $10000$ GRBs and with $f=1$ is small enough that there is a significant probability to observe no lensing events. 

The fractional abundance constraints have a typical behaviour where the constraints are tightest at a particular mass range (about $10^{-15} M_\odot$ for LEO) and get worse on either side of the mass range. This can be understood as a competition between two effects. As we consider higher masses, the Einstein radius of the PBH lens becomes greater than the detector separation. This implies that the relative differences in the magnification reduce and thus it reduces the lensing cross-section where the parallax can be observed. On the other hand, as we consider smaller masses, the Einstein radius of a PBH lens decreases which results in an increase in the relative size of the source with respect to Einstein radius. The finite source size effect makes it harder to detect microlensing, thus reducing the cross-section. In any case, given Daksha's energy range, the grey region denotes the mass range where the wave effects are expected to further significantly restrict our possibility of measuring lensing parallaxes. 

Our results can be compared with those presented in \citet{jung2020gamma} shown with the blue dot-dashed line for low Earth orbit. In their paper, \citet{jung2020gamma}, the simplistic detection condition of using fractional differences in the magnifications while computing the detectability of the lensing parallax leads to much stronger constraints. However, as we have shown, fractional differences in magnified fluxes is not a realistic way to assign detectability of the lensing parallax and implementing realistic detectability condition degrades the constraints by more than two orders of magnitude as shown in Fig.~\ref{fig:constraints}.

We also note further that each of the two satellites in a LEO cannot observe the entirety of the sky at the same time. The Earth occults particular line-of-sight directions as observed from each of the satellites, by an solid angle that depends upon the height of the satellite in its orbit. Furthermore, the detectors have to be switched off near the South Atlantic Anomaly, which further reduces the operation time during which both satellites can observe events, simultaneously. Thus the total number of GRBs with which the lensing parallax can be potentially observed is expected to be even smaller.

The abundance constraints we presented above are heavily dependent upon the size of the parallax, i.e. the difference in the impact parameter of the source compared to the position of the lens as seen from the two satellites. Increasing this parallax is expected to cause a further improvement in constraints. \citet{jung2020gamma} consider a baseline corresponding to Lagrange point 2 in their paper. We have also explored similar larger separations for the satellites. As an example, in Fig.~\ref{fig:constraints}, we show the expected constraints if we consider one of the satellites in orbit around the Earth, while the other one was in orbit around the Moon. We have used the detector characteristics of Daksha for this exercise. 

In this case, we observe a significant improvement in the possible constraints from the two satellites, where we can find meaningful constraints using lensing parallax on the abundance of PBH in the window of our interest. Potential non-detections of lensing parallax with 10000 events can now constrain the fraction of PBH in dark matter to a few per cent. The best detection sensitivity is for PBH masses of order $10^{-12}M_\odot$. This sweet spot is a balancing act between the same two mechanisms as we discussed earlier in this section, but this spot is at a higher mass end, due to the increased separation.

This alternative possibility of having a future Gamma ray detection mission is not very far out of the realm of possibilities. For example, NASA is currently considering a SmallSat mission called Moon Burst Energetics All-sky Monitor (MoonBEAM) \citep{2021AAS...23731502H}. The satellite is expected to have a cis-lunar orbit at the Lagrange point 3 of the Earth Moon system. In addition to the increased parallax, such a satellite will have a very small Earth occultation solid angle, and will not have to deal with the South Atlantic Anomaly. At this location, we expect the background to be higher than that expected for LEO as the Earth's magnetic field is no longer effective at such distances. However, the background is expected to be much more stable at such locations.

We compare the constraints on the abundance of PBH possible with an Earth Moon detector separation with existing constraints in Fig.~\ref{fig:constraints_comp}. The detection of 10000 common GRBs with satellites separated by the Earth Moon distance, can constrain the PBH abundances in the mass range $[10^{-15}-10^{-7}]M_\odot$. We mention the caveat that we have assumed the MoonBEAM detectors to be similar to that of Daksha. In the absence of similarities, we will have to cross-calibrate the two detectors, understand the differences in the sensitivity in the different energy ranges that they probe. Despite these caveats, this synergy could potentially constrain the PBH abundance in a parameter space, that has been notoriously difficult to constrain.

\section{Summary}
\label{sec:summary}

The statistics of the differential magnification observed due to the parallax of a putative primordial black hole with respect to a distant GRB as seen from two separate vantage points can be used to constrain the abundance of such black holes. We have studied the feasibility to carry out such a lensing parallax experiment. Our study can be summarized as follows:

\begin{itemize}
    \item We computed the optical depth of lensing parallax towards a GRB due to primordial black holes following a monochromatic mass spectrum. We use the redshift of the GRB, its time scale and source size, the properties of its energy spectrum, and the lensing impact parameters in order to compute the total number of photons that will be detected by the two satellites, and whether they can detect the lensing parallax given their sensitivity. We fold this information in the optical depth computation.
    \item We use the parameters of the population of GRBs detected by \swiftbat along with a distribution of their spectral parameters and fluxes as observed with detectors with sensitivities similar to those that are designed to be onboard the twin satellite mission Daksha.
    \item Accounting for such realistic detection condition and demanding at least a 5-$\sigma$ statistical significance in the difference of fluxes of GRBs as observed from the two satellites in LEO, we find that if 10000 GRBs are simultaneously observed by the Daksha satellites, the probability to see an event ranges in a decreasing manner from $80$ to $50$ per cent in the case monochromatic PBHs in the mass range $[10^{-15}-10^{-12}]M_\odot$ make up the entirety of dark matter.
    \item In the case of a non-detection, the constraints on PBH abundance, will not be able to conclusively rule out that the entirety of dark matter is made up of PBHs in this mass range.
    \item Alternatively, we propose to carry out such an experiment with a separation between satellites equal to that between the Earth and the Moon. In such a case, an experiment involving 10000 GRBs observed by both satellites could potentially rule out that the entirety of dark matter is made up of primordial black holes in the mass range $[10^{-15}, 10^{-7}] M_\odot$.
\end{itemize}

Although our study considered the specific case for the Daksha mission, the formalism presented in our study can be easily applied to any such twin missions. In future work we also plan to consider the GRBs detected by pairs of detectors in the inter-planetary network and understand its potential implications on the abundance of primordial black holes in the window of our interest. Understanding the cross-calibration between the heterogeneous detectors in the inter-planetary network will remain one of our biggest challenges.

\section*{ACKNOWLEDGEMENTS}

We thank the anonymous referee for comments which improved the clarity of the paper. We thank Dipankar Bhattacharya, Shabnam Iyyani, Alexander Kusenko and Navin Chaurasiya along with Sukanta Bose, Masamune Oguri and Anupreeta More for useful discussions on the project. SM thanks Nick Kaiser for his insightful comments regarding the role of finite source size effects and wave optics in the HSC microlensing constraints which prompted SM to explore this project. PG acknowledges financial support provided by the University Grants Commission (UGC) of India. She is also grateful to IUCAA for providing hospitable environment to students. We acknowledge the use of the high performance computing facility Pegasus at IUCAA for this work.

\section*{Data availability}
The data used in this paper is publicly available from the GRB catalogues whose links have been provided in the footnotes of the paper.

\bibliographystyle{mnras}
\bibliography{Citations}

\appendix

\section{Wave optics limit}
\label{app:GOL}
 \begin{figure*}
     \includegraphics[scale=1.2] {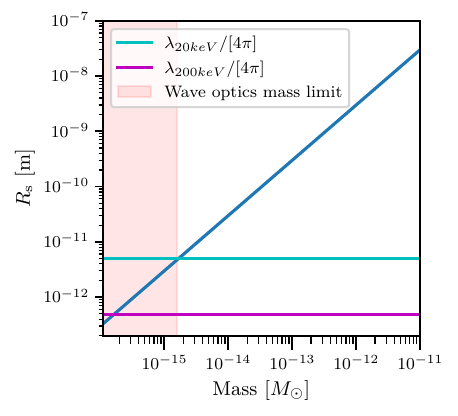} \caption{The Schwarzschild radius as a function of the mass of a PBH is shown using the blue solid line. We compare this to the wavelengths corresponding to the energy range of Daksha scaled by $4\pi$ as explained in Appendix~\ref{app:GOL}. A conservative use of $\zl = 0$, yields the shaded region to be governed by the wave optics limit for the energy range probed by Daksha.} \label{fig:Rsch} \end{figure*}
     
The condition for the applicability of geometric optics limit in gravitational lensing for light of frequency $f$ by a point mass $M$ is given by
\begin{align}
    2\pi f \td = 2 \pi \frac{c}{\lambda} \td \gg 1\,,
\end{align}
where $t_{\rm d}$ is the time delay between the various images \citep{Takahashi_2003}. The time delay depends upon the Fermat potential and is of the order of
\begin{align}
    \td = \frac{4GM}{c^3}(1+\zl)\,.
\end{align}
If we define, $r \equiv 4\pi(1+\zl)R_{\rm s}/\lambda$, where $R_{\rm s}$ is the Schwarzschild radius of the lens, then $r\gg1$ corresponds to the geometric optics limit. In this paper, we assume $r\ge 1$ to be sufficient for the validity of the geometric optics limit and show the comparison of $\lambda/(4\pi)$ with $R_{\rm s}$ in Figure~\ref{fig:Rsch}. In this case, ignoring the factor $(1+\zl)$, we obtain a limit of about $1.6\times 10^{-15} M_\odot$. Given the high median redshift of our GRB sources, the lensing efficiency kernel will peak around $\zl \sim 0.6$, which suggests the range of our validity of the geometric optics limit of $\sim 10^{-15}M_\odot$ and above. We note that changes to $\zl$ can change this range by no more than a factor of few, which is negligible compared to the large dynamic range in the mass range we consider in any of our figures. A stricter limit on $r$ for the validity of the geometric optics limit, could also alter the mass range for which our analysis is applicable.

\section{Finite source-size effect on the lensing cross section}
\label{app:finite}
The magnification of a finite sized circular source with constant surface brightness by a point-lens in the geometrical optics limit is given by \cite{1994ApJ...430..505W}
 \begin{equation} \label{eq:mag}
\mu(y, \delta) \approx \begin{cases}\mu_{\rm in}(y, \delta) & \text { for } y<\delta \\ \mu_{\rm out}(y, \delta)\,, & \text { for } y>\delta\end{cases}\,,
\end{equation}
where
\begin{equation}
    \begin{aligned}
\mu_{\rm in}(y, \delta)= & \sqrt{1+\frac{4}{\delta^2}}-\frac{8}{\delta^3\left(\delta^2+4\right)^{3 / 2}} \frac{y^2}{2} \\
& -\frac{144\left(\delta^4+2 \delta^2+2\right)}{\delta^5\left(\delta^2+4\right)^{7 / 2} }\frac{y^4}{24}, \,, \\
\mu_{\rm out }(y, \delta)= & \frac{2+y^2}{y \sqrt{y^2+4}}+\frac{8\left(y^2+1\right)}{y^3\left(y^2+4\right)^{5 / 2}} \frac{\delta^2}{2} \\
& +\frac{48\left(3 y^6+6 y^4+14 y^2+12\right)}{y^5\left(y^2+4\right)^{9 / 2}} \frac{\delta^4}{24}\,.
\end{aligned}
\end{equation}
Here, $\delta\equiv\left(r_{\rm s} / D_{\rm s}\right) / \theta_{\rm E}$ and $y\equiv\left(\beta/ \theta_{\rm E}\right)$. Around $y \approx \delta, \mu_{\text {in }}$ overestimates and $\mu_{\text {out }}$underestimates the magnification. In our calculation, we follow \citet{jung2020gamma} and linearly interpolate the magnification in the range $0.9 \delta<$ $y<1.1 \delta$.

\section{Cross-section in lensing parallax}
\label{app:parallax}

\begin{figure*}
     \includegraphics[scale=1.2] {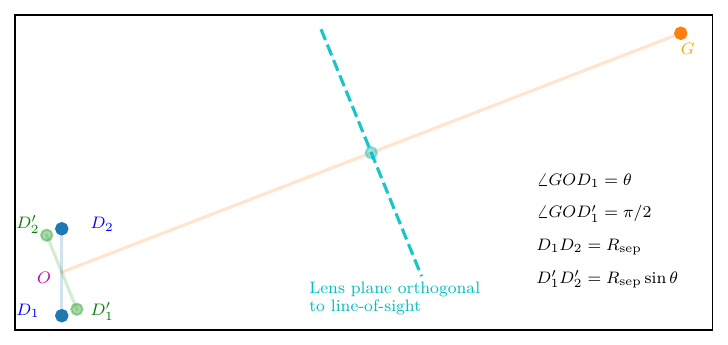} \caption{This illustrative figure shows the geometry we consider for computing the lensing parallax. The two satellites are marked as points $D_1$ and $D_2$. The line joining the two satellites makes an angle $\theta$ with respect to the line-of-sight towards the GRB marked as $G$. The separation relevant for the parallax calculation is the line segment $D_1'D_2'$ which is perpendicular to the line-of-sight to $G$. In order to compute the cross-section for lensing parallax in Appendix~\ref{app:parallax}, we consider the plane at the redshift of the lens that comes out of the paper and which is perpendicular to the line-of-sight. The dashed cyan line shows the $x_L$ or $x_L'$ axis which lies in the plane of the paper, and the $y_L$ or $y_L'$ axis corresponds to the normal to the plane of the paper.} \label{fig:example} \end{figure*}

\begin{figure}
     \includegraphics[scale=0.7] {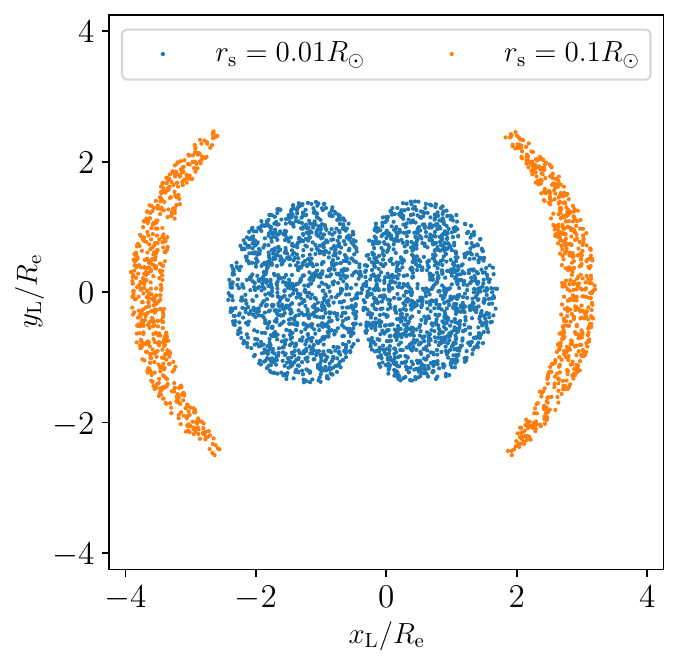} \caption{The potential locations of PBH lenses in the plane specified by the cyan dashed line in Fig.~\ref{fig:example} which lead to an appreciable lensing parallax observed at a significance larger than 5$\sigma$ for a GRB at $z=1$ with a fluence of $10^{-6}$ erg $\mathrm{cm}^{-2}$, and with a PBH lens of mass $10^{-15}M_\odot$ halfway between the GRB and the two satellites in LEO. The two colours depict the lensing cross-section for GRBs of different source sizes compared to the Einstein radius of the lens, the blue colour represents a small source size, while the orange colour shows the cross-section with a larger source size.} \label{lensing_crossection} \end{figure}

Consider a plane passing through the two satellites and the GRB source as shown in Fig.~\ref{fig:example}. The primordial black hole at the lens redshift defines a plane which is orthogonal to the line-of-sight i.e the line joining one of the satellites, $D_1$, and the source $G$. Without loss of generality we consider the intersection of these two planes, to be the $x_{\rm L}$-axis of our coordinate system and the orthogonal direction in the lens plane to be the $y_{\rm L}$-axis. Similarly, we can define $x'_{\rm L}-y'_{\rm L}$ coordinate system with respect to the line joining the satellite $D_2$ and the source. The origin of each of the coordinate systems is set to be the point where the line-of-sight to the source from either these satellites intersects the lens plane. The coordinates $y_{\rm L}$ and $y'_{\rm L}$ will be equal since they are out of the plane, while $x_{\rm L}$ and $x'_{\rm L}$ will be different due to the parallax angle expressed in Eq.~\ref{eq:parallax}.

We populate the $x_{\rm L}-y_{\rm L}$ plane with a number of points randomly drawn around the origin, with a distance 10 (25) times that of the Einstein radius for the case of low Earth Orbit (Earth-Moon case). We only consider those points which are closest to the satellite $D_1$ and discard those which are closer to $D_2$ to avoid double counting. We evaluate the magnification at each of these points as seen from both the satellites, and remove those points where the difference in the fluxes observed by the two satellites cannot be detected at more than 5 $\sigma$ given the sensitivity and the expected background noise. Two times the area corresponding to the points that survive represents the cross-section for the detection of lensing parallax.

We present example cross-sections for a PBH of mass $10^{-15}M_\odot$ located halfway between the two satellites and the GRB with a fluence of $10^{-6}\,{\rm erg cm^{-2}}$ at a redshift of unity in Fig.~\ref{lensing_crossection}. When the source size is small, the magnification is expected to rapidly fall off as the impact parameter of the lens from the true soure position increases. Locations near the lines-of-sight from the two satellites towards the GRB can contribute to appreciable lensing parallax, that allows for a detection in the difference in the GRB flux observed from the two vantage points. In these cases, the locations near the line-of-sight towards the GRB from, say one of the satellites, will have a larger magnification, while such positions will correspond to smaller magnification for the other satellite due to the larger impact parameter resulting from the parallax. Depending upon the size of the parallax, such separate regions around each of the satellites start to merge to form dumbbell shaped cross-sections as seen in the left hand panel of the Fig.~\ref{lensing_crossection}.

As we increase the source size, the magnification does not rise beyond a certain value even at decreasing impact parameters of the lens with respect to the source position. In such cases, depending upon the parallax, the lens positions which lie along the line-of-sight from each of the satellites towards true source positions, may not be sufficient to result in a detectable lensing parallax, as both source positions correspond to similar values of magnifications. In such cases, larger difference in the impact parameters is required to have appreciable lensing parallax and we start to see a crescent shape for the lensing cross-section, as seen in the right hand panel of Fig.~\ref{lensing_crossection}. In general, smaller source sizes are preferred for the lensing cross-section to be significant. However we have been also able to identify some cases where bigger source sizes in fact result in a larger cross-section, especially for sources that are bright enough and the parallax is large. This is a result of the finite source size effect. In addition to resulting in the magnification flattening out at very small impact parameters, the lensing magnification is larger at larger impact parameters than the point source size case. Thus it allows the lensing parallax to be detectable for large source size.

\bsp	
\label{lastpage}
\end{document}